\def\ba{\begin{eqnarray}}
\def\ea{\end{eqnarray}}
\def\nn{\nonumber \\}

\documentclass[12pt]{article}
 \begin{document} 
\title{Correlations and fluctuations: generalized factorial moments}

\author{A.Bialas \\ M.Smoluchowski Institute of Physics \\Jagellonian
University, Cracow\thanks{Address: Reymonta 4, 30-059 Krakow, Poland;
e-mail:bialas@th.if.uj.edu.pl;}}
\maketitle

PACS:


\begin{abstract}

A systematic study of the relations between fluctuations of the
extensive multiparticle variables and integrals of the inclusive
multiparticle densities is presented. The generalized factorial moments
are introduced and their physical meaning discussed. The effects of the 
additive conservation laws are analysed.

\end{abstract}
 
\section{Introduction}

Studies of multiparticle production is a complicated matter because of
the large number of variables describing the system and thus several
methods were developped to simplify the problem. Two of them turned out
particularly useful in providing information: (i) studies of
correlations between particles by investigation of the inclusive
distributions and (ii) studies of event-by-event fluctuations. Since
both of them give -in principle- a complete description of the system,
they must be related. This relation was firsst discussed in \cite{bk},
and reformulated in \cite{gfm}. In the present paper we follow the
arguments of \cite{bk,gfm} and investigate further the relation between
these two ways of approach to multiparticle production.

To illustrate the problem, consider the well-known formula
\begin{eqnarray}
F_k =\int dq_1 ...dq_k \rho_k(q_1,...,q_k)   \label{2.3a}
\ea
where $\rho_k(q_1,...,q_k)$ is the inclusive distribution of $k$
particles and $F_k$ is the factorial moment of the $k$-th order:
\begin{eqnarray}
F_k= \sum_N P(N) N(N-1)...(N-k+1) \equiv\left<N(N-1)...(N-k+1)\right> \label{fm}
\ea
 $N$ is the event multiplicity and  $P(N)$ denotes the multiplicity 
distribution.

The main interest in the formula (\ref{2.3a}) is that its L.H.S.
characterizes event-by-event fluctuations of the multiplicty while its
R.H.S. represents the measurement of particle distribution by $k$-arm
spectrometer. Thus (\ref{2.3a}) connects the two quantities which are
defined in an entirely different manner. Needless to say, confronting
such -apparently unrelated- quantities is often a source of a new
insight into the problem. This is precisely the interest in
investigation of further relations of this type.

Another feature of (\ref{2.3a}) which make this relation very useful in
practical applications is that it connects the factorial moment of order
$k$ with the inclusive density of the same order. Thus it can be used
even if one does not have a complete knowledge of the system (which is
of course practically always the case)\footnote{ This feature may be
contrasted with the relation between $F_k$ and $P(N)$. They must be
related because they both describe completely the multiplicity
distribution. However, as seen from (\ref{fm}), to calculate $F_k$ one
needs to know {\it all} $P(N)$ (and {\it vice versa}).}.

For $k=2$, the extension of (\ref{2.3a}) to quantities other than
multiplicity  was proposed in \cite{vs}. It was discussed and applied
by several authors (see, e.g., \cite{sv}-\cite{ag} and the review
\cite{jk}). The possibility of generalization to $k$ greater than $2$
was suggested in \cite{vs} and considered in \cite{v}. 

The purpose of the present paper is twofold. First, we derive an
explicit formula, valid for arbitrary $k$, which extends the the
relation (\ref{2.3a}) to fluctuations of quantities other than
multiplicity. This is obtained by introducing the {\it generalized}
factorial moments \cite{gfm}, defined in terms of measurements of
event-by-even fluctuations of measurable {\it extensive} quantities. Second,
we investigate the physical meaning of the generalized factorial moments
along the lines developped in \cite{bp1}, as described below.

As first discussed in \cite{bp1}, the factorial moments have a rather
simple physical interpretation.
 It follows from the 
observation that for any distribution of multiplicity which can be
represented as a superposition of Poisson distributions
\ba
P(N)=\int d\bar{N} W(\bar{N}) e^{-\bar{N}}\frac{\bar{N}^N}{N!}
\ea
one has
\ba
F_k=\sum_N (N(N-1)...(N-k+1) P(N)= \int d\bar{N} W(\bar{N})\bar{N}^k.
\label{px}
\ea
Thus the {\it factorial} moment of the actual multiplicity distribution
$P(N)$ measures the {\it standard} moment of the underlying distribution
$W(\bar{N})$. One can say \cite{bp1} that the factorial moment removes the
statistical noise (represented by the Poisson distribution) 
 from the corresponding moment of the distribution $W(\bar{N})$.
This observation is easily generalized to local multiplicity
fluctuations \cite{bp1}.

It is shown in the present paper that, in absence of other constraints,
the generalized factorial moments can be used as well to remove the
statistical noise from the data. This is of course to
be expected. The problem must be treated more carefully, however, in
presence of an additive conservation law. Indeed, the additive
conservation law implies that the "statistical noise" cannot be entirely
random and therefore the simple argument of \cite{bp1} must be modified.
We consider in some detail the effects of conservation of charge and of
transverse momentum.

The problem of removing the statistical noise from fluctuations of
various quantatities observed in multiparticle production, is the
long-standing one. The most popular method remains the $\Phi$ measure,
introduced in \cite{gm}. Other possibilities, and their relation to the 
$\Phi$ measure were discussed in \cite{vkr}-\cite{ag}. They 
are now commonly used in data analysis \cite{na49}-\cite{fbhb}. 

Our treatment, using the generalized factorial moments, goes beyond the
previous studies in two points. On the theoretical side, our method
applies to correlations of any order (until now mostly $k=2$ was
discussed\footnote{ A related method of removing the statistical noise,
applicable to arbitrary $k$, was described in \cite{fl} and studied in
\cite{jyj}.}). On the practical side, including the effects of additive
conservation laws makes the control over the reliabity of the
experimental results much more solid\footnote{For $k=2$ the effects of
charge conservation were discussed in \cite{pgv}. An interesting
proposal to deal with energy-momentum conservation can be found in
\cite{o}.}.

In the next section the generalized factorial moments are introduced and
their relation to integrals of inclusive multiparticle densities
explained. Their physical interpretation in case of random noise is
explained in Section 3. Modifications due to conservation of a discrete
quantum number are discussed on the example of charge distributions in
Section 4. The effects of conservation of transverse momentum are 
considered in Section 5. Discussion and conclusions are given in the
last section. Two Appendices explain the details of algebra.

\section {Generalized factorial moments}

Generalizations of (\ref{2.3a}) to fluctuations of extensive quantities
other than multiplicity were found in \cite{bk} and reformulated in
\cite{gfm}. Here we follow the argument used in these two papers.

Consider a single-particle variable $x=x(p)$, where $p$ is the particle
momentum. Consider, furthermore, the set of extensive quantities
\ba
X_l(q_1,...,q_N)=\sum_{n=1}^N [x(q_n)]^l   \label{q1}
\ea
where $N$ is the multiplicity of the event.

We want to study fluctuations of $X \equiv X_1$. One possible method 
is to consider the moments 
\ba
\left<X^k\right>= \sum_N\int dq_1...dq_N P(q_1,...,q_N;N) 
\left(x_1+...+x_N\right)^k        \label{q2}
\ea
where $P(q_1,...,q_N;N)$ is the probability to find an event with $N$
particles at momenta $(q_1,...,q_N)$ and where we have introduced the
shorthand
\ba
x_j\equiv x(q_j)   \label{q3}
\ea
which will be used henceforth. It was shown in \cite{bk} that the
moments (\ref{q2}) can be expressed by linear combinations of the
integrals
\ba
R(k_1,...,k_s;s)=\int dq_1...dq_s \rho_s(q_1,...,q_s)
x_1^{k_1}...x_s^{k_s}.
\ea
These relations are, however, fairly complicated. 

To find a more elegant formulation we observe that when one takes $x\equiv
1$, one has $X\equiv  N$ and thus the simple relation (\ref{2.3a}) must hold.
It is thus clear that one has to find a generalization of the factorial
moments (\ref{fm}). Such a generalization was proposed in 
\cite{gfm}:
\ba
F_k[x]= \left<[X-(k-1)\hat{x}][X-(k-2)\hat{x}]...[X-\hat{x}]X\right> \label{q4}
\ea
where $\hat{x}$ is the operator, acting on a product $[X_{l_1}...X_{l_m}]$ 
as follows\footnote{The
definition of the operator $\hat{x}$ given in \cite{gfm} was incomplete.
I would like to thank Andrzej Kotanski for pointing out the error.}
\ba
\hat{x} [X_{l_1}...X_{l_m}]
=\frac1{m}\sum_{s=1}^m[X_{l_1}...X_{l_s+1}...X_{l_m}]  \label{q4c}
\ea
where the indices $[l_1,...,l_m]$ need not be different.
It follows from (\ref{q4c}) that, in particular,
\ba
\hat{x} X_l=X_{l+1}\;;\;\; \hat{x}[X_l]^m=X_{l+1}[X_l]^{m-1}
\ea

From (\ref{q4c}) one can also deduce how $\hat{x}$ acts on a product
$x_{j_1}...x_{j_k}$. The result is 
\ba
\hat{x}[x_{j_1}...x_{j_k}]=(x_{j_1}...x_{j_k})\frac1{k}
\sum_{s=1}^kx_{j_s}  \label{q4a}
\ea
where, again,  the indices $j_1,...,j_k$ need not be different.

Using (\ref{q4}) and (\ref{q4c}) we obtain for $k=1,2,3$ 
\ba
F_1[X]= \left<X\right>\;;\;\; 
F_2[X]=\left<X^2\right>-\left<\hat{x}X\right>=
\left<X^2\right>-\left<X_2\right>\;;\nonumber \\ F_3[X]=
\left<X^3-2\hat{x}X^2-X\hat{x}X+2[\hat{x}]^2X\right>=
\left<X^3-3XX_2+2X_3\right>\;.
\ea

We shall now prove that 
\ba
[X-(k-1)\hat{x}][X-(k-2)\hat{x}]...[X-\hat{x}]X=\sum_{i_1=1}^N...
\sum_{i_k=1}^N x_{i_1}...x_{i_k}  \label{q4b}
\ea
where all indices are different from each other.

The proof goes by induction. For $k=2$ we have $X^2- X_2 = \sum_{i=1}^N
\sum_{j=1}^N x_ix_j$ with $i\neq j$. Thus (\ref{q4b}) is satisfied.
Suppose now that it is valid for a given $k$. Multiplying both sides of
(\ref{q4b}) by $(X-k\hat{x})$ we thus have
\ba
[X-k\hat{x}][X-(k-1)\hat{x}]...[X-\hat{x}]X=(X-k\hat{x})\sum_{i_1=1}^N...
\sum_{i_k=1}^N x_{1_1}...x_{i_k}  \label{q8}
\ea
But
\ba
X\sum_{i_1=1}^N...\sum_{i_k=1}^N x_{1_1}...x_{i_k}=\sum_{j=1}^N x_j
\sum_{i_1=1}^N...\sum_{i_k=1}^N x_{1_1}...x_{i_k}=\nonumber \\=
\sum_{s=1}^k\sum_{i_1=1}^N...\sum_{i_k=1}^N x_{1_1}...[x_{i_s}]^2...x_{i_k}
+\sum_{s=k+1}^N\sum_{i_1=1}^N...\sum_{i_k=1}^N
x_{1_1}......x_{i_k}x_{i_s} \label{q6}
\ea
where in the first term in the R.H.S. 
one of the variables from $X$ is identical to
one of $x_{i_1}...x_{i_k}$, whereas in the second term all variables
differ from each other. Noting that, as seen from (\ref{q4a}), the first
term is identical to 
\ba
k\hat{x}[\sum_{i_1=1}^N... \sum_{i_k=1}^N
x_{1_1}...x_{i_k}],
\ea
 one sees that it cancels in (\ref{q8}) and thus
R.H.S. of (\ref{q8}) equals to the second term in the R.H.S. of
(\ref{q6}). But this term is just what is needed to complete the proof. 

It follows from (\ref{q4b}) that 
\ba
F_k[X]\equiv \sum_{N=0}^\infty \int dq_1...dq_N P(q_1,...,q_N;N)
[X-(k-1)\hat{x}]...[X-\hat{x}]X=\nonumber \\=
\int \rho_k(q_1,...,q_k) x(q_1)...x(q_k) dq_1...dq_k  \label{q5}
\ea
where $\rho_k(q_1,...,q_k)$ is the $k$-particle inclusive density. This
follows almost directly from definition of inclusive density
\ba
\rho_k(q_1,...,q_k)= \sum_N N(N-1)...(N-k+1) \int dq_{k+1}...dq_N
P(q_1,...,q_N;N)   \label{qa5}
\ea
if one notices that the number of terms in the R.H.S. of (\ref{q4b})
equals $N(N-1)...(N-k+1)$. 

Eq. (\ref{q5}) represents a generalization of (\ref{2.3a}) (which is
obtained from (\ref{q5}) by putting $x(q)\equiv 1$). It has the same two
attractive features: it connects the fluctuations of the extensive
variables $X_l$ with the integral of the inclusive density and it
involves only moments of finite order.

\section {Physical interpretation of the generalized factorial moments}

To discuss the physical interpretation of the generalized factorial
moments, we again consider the momentum space split into s-bin and the
rest. To determine the generalized factorial moment in the s-bin, we
need the distribution of the number of particles and of the variable $x$
in it. Following the lines of \cite{bp1}, we demand that in each
bin particles are distributed as randomly as possible. In absence of any
constraints this implies in the s-bin  the distribution 

\ba
p_s(q_{1},...,q_{n},n;\bar{n},\eta)dq_{1}...dq_{n}=
e^{-\bar{n}}\frac{[\bar{n}]^{n}}{n!}
\prod_{l=1}^{n}\left[g(q_{l};\eta)dq_{l}\right]   \label{p3}
\ea
where $\bar{n}$ is the average multiplicity in the s-bin  and 
$g(q;\eta)$ is the momentum distribution of one particle on this bin, with
$\eta$ representing a collection of parameters on which this
distribution may depend.

To obtain the actual distribution  we have to weight
(\ref{p3}) by the underlying probability distribution
$W_s(\bar{n},\eta)$:
\ba
P_s[q_{1},...,q_{n},n]
=\int d\bar{n} d\eta_m W_s(\bar{n},\eta)
p_s(q_{1},...,q_{n},n;\bar{n},\eta) .  \label{p5}
\ea

From (\ref{p5}) we deduce the inclusive particle density in the s-bin 
\ba
\rho_k(q_{1},...q_{k})=\sum_{n_s} n_s(n_s-1)...(n_s-k+1)
\int dq_{k+1}...dq_{n} P_s(q_{1},...,q_{n};n)  \label{p6}
\ea
Introducing   (\ref{p6}) into (\ref{q5}) and using (\ref{p3})  we obtain
 for the generalized factorial moment in the s-bin
\ba
F_k[X]= \sum_{n} n(n-1)...(n-k+1)
\int dq_{1}...dq_{n} P_s(q_{1},...,q_{n};n)x( q_{1})...x(q_{k})
\nonumber \\= \int d\bar{n} d\eta  W_s(\bar{n},\eta)
[\bar{n}\bar{x}]^k \equiv \left<[\bar{n}\bar{x}]^k\right> =
\left<\bar{X}^k\right> \label{p9}
\ea
where 
\ba
\bar{x}=\bar{x}(\eta)= \int  g(q;\eta) x(q)dq
\ea
is the average  of $x(q)$   at fixed $\eta$. 
$\bar{X}$ is the average of $X$ at fixed $\bar{n}$ and $\eta$.

Eq. (\ref{p9}) is the generalization we were seeking for. It shows the
physical interpretation of the generalized factorial moments: they
remove the statistical noise from the moments of
$\bar{n}\bar{x}=\bar{X}$ of the underlying distribution $W_s$. For
$x\equiv 1$ we of course recover the well-known result \cite{bp1}.

\section {Fluctuations of charge}

As already explained in the introduction, the argument present in the
previous section fails in case when the variable $x$ satisfies an
additive conservation law. In this section we discuss fluctuations of
charge. The discussion applies to any discrete, additive quantum number.

We again select an s-bin from the momentum space and consider the
distribution of particles inside and outside of this bin.

The first problem is to define the distribution describing the
"statitical noise". Following the idea of \cite{bp1}, we take the
normalized product of Poisson distributions (expressing the independent
particle emission and thus introducing no correlations). This is
supplemented by the Kronecker $\delta$-symbol to satisfy the
conservation law:
\ba
p_s=\frac1{A_Q}
e^{-\hat{n}^+}\frac{[\hat{n}^+]^{n^+}}{n^+!}
e^{-\hat{n}^-}\frac{[\hat{n}^-]^{n^-}}{n^-!}
e^{-\hat{N}^{*+}}\frac{[\hat{N}^{*+}]^{N^{*+}}}{N^{*+}!}
e^{-\hat{N}^{*-}}\frac{[\hat{N}^{*-}]^{N^{*-}}}{N^{*-}!}\nonumber \\
\delta_{n^++N^{*+}-n^--N^{*-}-Q_0} \label{l1} 
\ea 
where $n^+,n^-$ denote
the number of produced  charges in the s-bin, $
N^{*+},N^{*-}$ those outside of the s-bin.  $Q_0$ is the
net charge in the initial state. We take $Q_0\geq 0$. 
The normalization factor 
$A_Q$ is given by (see Appendix 1)
\ba
A_Q = e^{-\hat{N}}\omega^{-Q_0} I_{Q_0}(\tilde{N})   \label{lx1}
\ea
where $I_{Q_0}$ denotes the Bessel function, 
\ba
 \hat{N} =\hat{n}^++\hat{N}^{*+}+\hat{n}^-+\hat{N}^{*-} \equiv 
\hat{N}^++\hat{N}^- \label{y1}
\ea
 and
\ba
\tilde{N}=
2\sqrt{\hat{N}^+\hat{N}^-}\;;\;\;\omega=\frac{2\hat{N}^-}{\tilde{N}}=
\frac{\tilde{N}}{2\hat{N}^+}   \label{df}
\ea

The observed distribution of particles is thus
 \ba
P(n^+,n^-,N^{*+},N^{*-};Q_0)=\int
d\hat{n}^+d\hat{n}^-d\hat{N}^{*+}d\hat{N}^{*-}
W(\hat{n}^+,\hat{n}^-;\hat{N}^{*+},\hat{N}^{*-})\nonumber \\
p_s(n^+,n^-;N^{*+},N^{*-};Q_0;
\hat{n}^+,\hat{n}^-;\hat{N}^{*+},\hat{N}^{*-}) \label{l01} 
\ea 
where $W$ is the "dynamical" distribution, free of the statistical
noise\footnote{The selection of statistical noise in the form of
(\ref{l1}) is not unique. Another natural possibility is to consider
independent emission of particle pairs (supplemented by independent
emission of $Q_0$ positive particles). The two methods differ mostly by
the treatment of the initial charge $Q_0$. Without further information
about dynamics it is difficult to decide which of them describes better
the physics of the problem.}.

It is important to observe that, contrary to  the naive expectation,
the parameters $\hat{n}^+,\hat{n}^-; \hat{N}^{*+},\hat{N}^{*-}$ 
 do not represent the average values
of $n^+,n^-;\\ N^{*+},N^{*-}$, respectively. 
The explicit formulae for these average values are given in Appendix 1.
In particular,
 the average values of the number of
produced positive and negative particles in the the s-bin are given by
\ba
\left<\bar{n}^\pm\right>= \left<\hat{n}^\pm \omega^{\pm 1}
\frac{I_{Q_0\mp 1}(\tilde{N})}{I_{Q_0}(\tilde{N})}\right>=
 \left<\frac{\hat{n}^\pm}{\hat{N}^\pm}\frac{\tilde{N}}2
\frac{I_{Q_0\mp 1}(\tilde{N})}{I_{Q_0}(\tilde{N})}\right> , \label{af}
\ea
where we have denoted by $\bar{}$ the average at fixed 
$\hat{n}^+,\hat{n}^-;\hat{N}^{*+},\hat{N}^{*-}$,
 and by  $\left<...\right>$ the average over 
$\hat{n}^+,\hat{n}^-;\hat{N}^{*+},\hat{N}^{*-}$  (with the probability 
distribution $W$).

The explicit formulae for the generalized factorial
moments of the charge $F_k^-$ and the (standard) factorial moments of
the multiplicity $F_k^+$ in the s-bin are derived in  Appendix 1. They read:
\ba
F_k^\pm=
\left<\left(\frac{\tilde{N}}2\right)^k\sum_{m=0}^k \frac{k!}{(k-m)!m!}
\left[\frac{\hat{n}^+}{\hat{N}^+}\right]^{k-m}
\left[\pm \frac{\hat{n}^-}{\hat{N}^-}\right]^m
 \frac{I_{Q_0+2m-k}(\tilde{N})}
{I_{Q_0}(\tilde{N})}\right> \label{final}
\ea

Eqs. (\ref{af}) and (\ref{final}) simplify substantially in the
interesting limit of very large number of produced particles,
$ \bar{N}\approx\tilde{N}\;\rightarrow\;\infty$.
 In this
limit, using the asymptotic expansion of the two Bessel functions, one
obtains up to the terms of order $\bar{N}^{-1}$ 
\ba F_k^\pm=\left<[\bar{n}^+\pm \bar{n}^-]^k
-\frac{k(k-1)}{2\bar{N}}[\bar{n}^+\pm\bar{n}^-]^{k-2}
\left[\bar{n}^+\mp\bar{n}^-\right]^2\right>. \label{ffox}
 \ea

The first term corresponds to the standard interpretation of the
generalized factorial moments, as expressed by (\ref{p9}). One sees from
(\ref{ffox}) that the correction to this result (induced by the
conservation law) vanishes, at fixed $\bar{n}^\pm$, with the inverse
power of the total multiplicity of produced particles. In this case the
correction is expected to be small at high energies (particularly for
the central heavy ion collisions)\footnote{For $F_k^-$ the {\it
relative} correction may be large if $\bar{n}_++\bar{n}_- \gg
\bar{n}_+-\bar{n}_-$.}. We thus conclude that the generalized factorial
moments can indeed provide a useful tool for eliminating the statistical
noise from the event-by-event fluctuations of multiplicity ($F_k^+$) and
charge ($F_k^-$) even in presence of the conservation law.

\section { Fluctuations in transverse momentum}

In this section we discuss fluctuations of the tranverse momentum, as an
example of a continuous variable subject to an additive conservation
law.

Selecting one bin in rapidity (s-bin) and dividing the available
momentum phase-space into this bin and the rest, we  write the
transverse momentum distribution in the form
\ba
P[q_{1},...,q_{n},n; q^*_1,...,q_{N^*}^*,N^*;q_0]=\nonumber \\
=\int d\hat{n} d\hat{q}dD d\hat{N}^* d\hat{q}^* dD^*
W(\hat{n} ,\hat{q},d, ;\hat{N}^* ,\hat{q}^*, D^*)\nonumber\\
e^{-\hat{n}}\frac{[\hat{n}]^{n}}{n!}
\prod_{l=1}^{n}\left[ \frac{e^{-(q_{l}-\hat{q})^2/2D^2}}{2\pi D^2} 
\right]
e^{-\hat{N^*}}\frac{[\hat{N}^*]^{N^*}}{N^*!}
\prod_{l=1}^{N^*}\left[ \frac{e^{-(q_{l}^*-\hat{q}^*)^2/2(D^*)^2}}
{2\pi (D^*)^2}  \right] \nonumber \\
\frac1{A_\perp}\delta(q_{1}+...+q_{n}+ q^*_1+...+q_{N^*}^*-q_0) \label{t2}
\ea
where $q$ denotes the two-dimensional transverse momentum vector and
$q_0$ is the total transverse momentum of the considered phase-space
region. The uncorrelated "statistical noise" in the form of Poisson
distribution for multiplicity multiplied by a gaussian distribution of
transverse momenta is supplemented by the $\delta$ function insuring the
transverse momentum conservation\footnote{ This is a commonly used form
of the "statistical noise" (see, e.g., \cite{o,bdo} ). Note that, since
$W(\hat{n} ,\hat{q},d, ;\hat{N}^* ,\hat{q}^*, D^*)$ is a general
(positive) function, the assumed gaussian forms in (\ref{t2}) do not
restrict seriously the observed distribution of particles (as both the
average values, $\hat{q}, \hat{q}^*$ and dispersions, $D, D^*$ of the
Gaussians can fluctuate according to the distribution $W$).} and by the
normalization factor $A_\perp$.

It is also natural to  take
\ba
\hat{n}\hat{q}+\hat{N}^*\hat{q}^*  =q_0 . \label{o3}
\ea

Replacing the $\delta$ function in (\ref{t2}) by its Fourier transform
we obtain for the inclusive distribution in the s-bin 
\ba
\rho_k(q_{1},...q_{k})=
\int  d\hat{N} d\hat{q} d Dd\hat{N}^* d\hat{q}^* d D^*
W(\hat{n}, \hat{q} , D;\hat{N}^*, \hat{q}^* , D^*) \nonumber \\
e^{\hat{N}^*[e^{i\hat{q}^*y-D^{*2}y^2/2}-1]}
\frac1{(2\pi)^2A_\perp} \int dy e^{-iyq_0}
\sum_{n} n(n-1)...(n-k+1)
e^{-\hat{n}}\frac{[\hat{n}]^{n}}{n!} \nonumber \\
\int dq_{(k+1)}...dq_{n} 
\prod_{l=1}^{n}\left[ \frac1{2\pi D^2}e^{iq_ly} e^{-(q_{l}-\hat{q})^2/2D^2}
\right]
  \label{t4}
\ea

Using (\ref{q5}) we thus have for the generalized factorial moment in the 
 bin $s$ 
\ba
F_k[X;q_0] =\int d\hat{n}d\hat{q}
 d D d\hat{N}^* d\hat{q}^* d D^*
W(\hat{n}, \hat{q} , D;\hat{N}^*, \hat{q}^* , D^*) \frac{S_{k\perp}}
{S_{0\perp}} \equiv \left<\frac{S_{k\perp}}
{S_{0\perp}}\right>
\label{t6a}
\ea
with 
\ba
S_{k\perp}=\frac1{(2\pi)^2}\int dy e^{-iyq_0}\left[\hat{n} \phi(y)\right]^k
e^{\hat{n}[e^{i\hat{q}y-D^{2}y^2/2}-1]}
e^{\hat{N}^*[e^{i\hat{q}^*y-D^{*2}y^2/2}-1]}    \label{t7}
\ea
(note that $A_\perp= S_{0\perp}$), and
\ba 
\phi(y) =\frac1{2\pi D^2}
\int d^2q e^{-(q-\hat{q})^2/2D^2} e^{iqy}x(q)\;;\;\;
\phi(0)=\bar{x}\label{t7a}
\ea
where $\bar{x}$ is the average value of $x$ at fixed $\hat{q}$ and
$D^2$.  
The function $\phi(y)$ depends obviously on the choice of the variable
$x(q)$.  Generally, if $x(q)$ is a
polynomial in $q$ of order $l$, then $\phi(y)$ is a polynomial in $y$ of
the same order\footnote{E.g., for $x(q)\equiv q$ we have $\phi(y)= (\hat{q}+iyD^2)
e^{i\hat{q} y -y^2D^2/2}$, for $x(q)=q^2$ we obtain $\phi(y)=
[2D^2+(\hat{q}+iyD^2)^2]e^{i\hat{q} y -y^2D^2/2}$.}
 multiplied by $e^{i\hat{q} y -y^2D^2/2}$.

If the number of particles is  large, the integral over $d^2y$ can be 
evaluated by the saddle point method. We write
\ba
S_{k\perp}=\frac1{(2\pi)^2}\int d^2y \left[\hat{n} \phi(y)\right]^k
e^{\Psi(y)}              \label{t10}
\ea
with
\ba
\Psi(y)= -iq_0y  +
{\hat{n}[e^{i\hat{q}y-D^{2}y^2/2}-1]}+
{\hat{N}^*[e^{i\hat{q}^*y-D^{*2}y^2/2}-1]} \label{t11}
\ea
The saddle point equation $\Psi'=0$ gives $y_0 \sim
q_0-\hat{n}\hat{q}-(\hat{N}^*\hat{q}^*)$ and thus (\ref{o3}) implies
$y_0=0$.
Furthermore $\Psi''(0)=-\hat{N}(\tilde{D}^2 +\tilde{q}^2)\equiv -\hat{N}\Delta^2$, 
where $\hat{N}=\hat{n}+\hat{N}^*$ and 
\ba
\tilde{D}^2= \frac{\hat{n} D^2 +\hat{N}^* D^{*2}}{\hat{N}}\;;\;\;
\tilde{q}^2= \frac{\hat{n}\hat{q}^2+\hat{N}^*\hat{q}^{*2}}
{\hat{N}} \label{t13}
\ea

Consequently we obtain
\ba
\frac{S_{k\perp}}{S_{0\perp}}\approx 
\frac{\hat{N}\Delta^2}{2\pi}\int d^2y \left[\hat{n} \phi(y)\right]^k
e^{-y^2\hat{N}\Delta^2/2 } .  \label{t16}
\ea

To evaluate this integral in the limit of large $\hat{N}$, one can
expand $\phi(y)$ in powers of $y$ and observe that, since the
coefficient in the exponent increases with increasing $\hat{N}$, the
dominant term at large $\hat{N}$ is that with the lowest power of $y^2$
under the integral (the terms with odd powers of $y$ do not contribute).
A detailed discussion is given in Appendix 2.  Here we anly summarize
the results.

If $\phi(0) =\bar{x} \neq 0$ one obtains
\ba
F_k[X]= \left<[\bar{n}\bar{q}]^k\right>
\ea
i.e. we recover the formula (\ref{p9}). Thus, if the average value of
$x(q)$ in the $s$-bin does not vanish, the standard interpretation 
of the generalized factorial moments remains valid, even in presence of
the conservation law. The corrections to this result vanish as 
$1/<N>$. They are given in (\ref{t16y}) of Appendix 2.

This is not the case if $\bar{x} =0$ and $\overline{qx}\neq 0$. For $k$
even, $k=2p$, one obtains 
\ba F_{2p}[X]= (-1)^p
p!\left<[\bar{n}\overline{qx}]^p \left(\frac{2\bar{n}\overline{qx}}
{\hat{N}\Delta^2}\right)^p \right> \label{tp} 
\ea 
which shows that now
the factorial moments are related to the average value $\overline{qx}$
rather than to $\bar{x}$. For a fixed $\bar{n}$ and large $\bar{N}$
these moments tend to zero. For a finite ratio $\bar{n}/\bar{N}$,
however, they may be large. The corrections to (\ref{tp}) and the
 formula for $k$ odd are given in Appendix 2.

\section {Discussion}

We have reconsidered relations between the event-by-event fluctuations of
 extensive multiparticle variables and the integrals of the inclusive
distributions \cite{bk}. Our main result is the formula (\ref{q5}) which
expresses the integral of the $k$-particle inclusive distribution in
terms of the generalized factorial moment of the $k$-th order. 

The physical meaning of the generalized factorial moments is discussed.
It is shown that they remove the random uncorrelated statistical noise
from the data, the result known already from previous investigations
\cite{bp1,gfm}. When an additive conservation law is at work, however,
the statistical noise cannot be uncorrelated and, consequently,  its
removal can at best be approximate. The exact formulae were derived for
charge and transverse momentum conservation. The corrections were
evaluated in the limit of very large total number of produced particles,
relevant for collisions at high incident energy.

In short, the results presented in this paper formulate a systematic
approach to investigation of fluctuations of extensive variables. They
seem to be particularly useful for studies aiming to uncover the {\it
local} structure of the multiparticle system. To take just one example,
interpretation of the data on transverse momenta (or energies) in small
rapidity bins will be much more transparent if presented in terms of the
generalized factorial moments. 

It was shown in \cite{bk} that the relation between fluctuations and
correlations can be extended to other extensive variables. This,
however, goes beyond the scope of the present investigation.

\section {Appendix 1}

Replacing the Kronecker $\delta$ symbol  by its integral representation
one obtains
\ba
P(n^+,n^-;N^{*+},N^{*-};Q_0)=\left<
w_Q(n^+,n^-;N^{*+},N^{*-};Q_0)\right>
\ea
where
\ba
w_Q(n^+,n^-;N^{*+},N^{*-};Q_0)=\frac1{ A_Q}
\frac1{2\pi}\int_0^{2\pi} df e^{-iQ_0f}\nonumber \\ e^{-\hat{n}^+}\frac{[\hat{n}^+e^{if}]^{n^+}}{n^+!}
e^{-\hat{n}^-}\frac{[\hat{n}^-e^{-if}]^{n^-}}{n^-!}
e^{-\hat{N}^{*+}}\frac{[\hat{N}^{*+}e^{if}]^{N^{*+}}}{N^{*+}!}
e^{-\hat{N}^{*-}}\frac{[\hat{N}^{*-}e^{-if}]^{N^{*-}}}{N^{*-}!}
\label{l3}
\ea
and $A_Q$ is the normalization factor.

Thus the generating function for the distribution $w_Q$ is
\ba
\Phi_Q(z^+,z^-,Z^+, Z^- )\equiv\nn
\sum_{n^\pm,N^{\pm*}}w_Q(n^+,n^-;N^{*+},N^{*-};Q_0)
(z^+)^{n^+}(z^-)^{n^-}(Z^+)^{N^{+*}}(Z^-)^{N^{-*}}
=\nn=
\frac1{2\pi}\int_0^{2\pi} df e^{-iQ_0f}
e^{\hat{n}^{+}(z^+e^{if}-1)}e^{\hat{n}^{-}(z^-e^{-if}-1)}
e^{\hat{N}^{+*}(Z^+e^{if}-1)}e^{\hat{N}^{-*}(Z^-e^{-if}-1)}
\ea
From this it is not difficult to derive, by the standard methods,
  the average multiplicities. One obtains
\ba
\bar{n}^\pm= \frac1{2\pi A_Q}\int_0^{2\pi} df e^{-iQ_0f}
e^{\hat{N}^{+}(e^{if}-1)}e^{\hat{N}^{-}(e^{-if}-1)}
\hat{n}^\pm e^{\pm if}=\nonumber \\=
\frac{e^{-\hat{N}}}
{2\pi i A_Q}\oint \frac{dz}{z}e^{\tilde{N}(z+1/z)/2}(\omega z)^{-Q_0}
 \hat{n}^\pm(\omega z)^{\pm 1}=
\nonumber \\=
\frac{e^{-\hat{N}}}{A_Q}
 \hat{n}^{\pm 1}\omega^{\pm1-Q_0}
\sum_{j=-\infty}^\infty I_j(\tilde{N})\frac1{2\pi i} \oint \frac{dz}{z} 
z^j z^{\pm1-Q_0}=\nonumber \\=
 \hat{n}^\pm\omega^{\pm1}\frac{I_{Q_0\mp1}(\tilde{N})}{I_{Q_0}(\tilde{N})}
  \label{finales}
\ea
where $\omega$ and $\tilde{N}$ are defined in (\ref{df}).

Similarly, one obtains
\ba
\bar{N}^\pm = \hat{N}^\pm\omega^{\pm1}
\frac{I_{Q_0\mp1}(\tilde{N})}{I_{Q_0}(\tilde{N})}.
\ea
It follows that
\ba
\hat{n}^\pm\omega^{\pm 1} =\frac{\tilde{N}}2 \frac{\hat{n}^\pm}{\hat{N}^\pm}=
\frac{\tilde{N}}2\frac{\bar{n}^\pm}{\bar{N}^\pm}  \label{barhat}
\ea

The inclusive distribution in the bin $s$ is
\ba
\rho_{k}(k^+,k^-)= \left<\frac1{2\pi A_Q }
\int_0^{2\pi} df e^{-iQ_0f} H^Q(k^+,k^-;f)\right> \label{14b}
\ea
where
\ba
H^Q(k^+,k^-,f)= e^{\hat{N}^{*+}(e^{if}-1)}
e^{\hat{N}^{*-}(e^{-if}-1)}
\sum_{n^+,n^-} V(n^+,n^-;k^+,k^-)
\nonumber \\
e^{-\hat{n}^+}\frac{[\hat{n}^+e^{if}]^{n^+}}{n^+!}
e^{-\hat{n}^-}\frac{[\hat{n}^-e^{-if}]^{n^-}}{n^-!}
\label{l3a}
\ea
and 
\ba
V(n^+,n^-;k^+,k^-)=\delta_{k-k^+-k^-}\frac{k!}{k^+!k^-!}
\frac{n^+!}{(n^+-k^+)!}
\frac{n^-!}{(n^--k^-)!}=\nonumber \\=\frac1{2\pi}\int_0^{2\pi} dh e^{-ihk}
k!\frac{n^+!}{k^+!(n^+-k^+)!} e^{ihk^+}
\frac{n^-!}{k^-!(n^--k^-)!}e^{ihk^-}
\ea
is the number of ways one can select $k^+$ out of $n^+$ positive and 
$k^-$ out of $n^-$ negative particles in a given order. 

To calculate the generalized factorial moment $F_k^-$ in
 the s-bin 
\ba
F_k^-=\sum_{k^++k^-=k} (-1)^{k^-}\rho_k(k^+,k^-)  \label{fact}
\ea
we observe that
\ba
\sum_{k^+}\sum_{k_-}(-1)^{k^-} V(n^+,n^-;k^+,k^-)=
\frac{k!}{2\pi}\int_0^{2\pi} dh e^{-ihk}[1+e^{ih}]^{n^+}[1-e^{ih}]^{n^-}
\ea

and thus
\ba
 \sum_{k_+,k_-}(-1)^{k^-}H^Q(k^+,k^-,f)=\nonumber \\
e^{\hat{N}^{+}(e^{if}-1)}e^{\hat{N}^{-}(e^{-if}-1)}
\frac{k!}{2\pi}\int_0^{2\pi} dh e^{-ihk}
e^{[\hat{n}^+e^{if}-\hat{n}^-e^{-if}]e^{ih}}=\nonumber  \\=
e^{\hat{N}^{+}(e^{if}-1)}e^{\hat{N}^{-}(e^{-if}-1)}
[\hat{n}^+e^{if}-\hat{n}^-e^{-if}]^k     \label{l3ab}
\ea

Now integration over $df$:
\ba
\frac1{2\pi }\int_0^{2\pi} df e^{-iQ_0f}\sum_{k^++k^-=k} H^Q(k^+,k^-;f)=
 \nonumber \\=
\frac1{2\pi}\int_0^{2\pi} df e^{-iQ_0f}
e^{\hat{N}^{+}(e^{if}-1)}e^{\hat{N}^{-}(e^{-if}-1)}
[\hat{n}^+e^{if}-\hat{n}^-e^{-if}]^k=\nonumber \\=
\frac{e^{-\hat{N}}}
{2\pi i}\oint \frac{dz}{z}e^{\tilde{N}(z+1/z)/2}(\omega z)^{-Q_0}
 [\hat{n}^+\omega z-\hat{n}^-/(\omega z)]^k=\nonumber \\=
e^{-\hat{N}}\omega^{-Q_0}\sum_{m=0}^k \frac{k!}{(k-m)!m!}
[-\hat{n}^-/\omega]^m[\hat{n}^+\omega]^{k-m} \nonumber \\
\sum_{j=-\infty}^\infty I_j(\tilde{N})\frac1{2\pi i} \oint \frac{dz}{z} 
z^j z^{k-2m-Q_0}=\nonumber \\=
e^{-\hat{N}}\omega^{-Q_0}\sum_{m=0}^k \frac{k!}{(k-m)!m!}
[-\hat{n}^-/\omega]^m[\omega\hat{n}^+]^{k-m} I_{Q_0+2m-k}(\tilde{N}),
  \label{finale}
\ea
 where $\hat{N}^\pm, \tilde{N}$ and $\omega$ are defined in 
Eqs. (\ref{y1}) and (\ref{df}).

The normalization factor $A_Q$ is obtained from (\ref{finale}) by putting
$k=0$. Inserting (\ref{finale}) and (\ref{barhat}) into (\ref{14b}) and (\ref{fact})
 we obtain (\ref{final}). 

We also note  that the standard factorial moment for the
multiplicity can be obtained from (\ref{finale}) simply by replacing 
$[-\hat{n}^-]^m$ by $[\hat{n}^-]^m$.

\section {Appendix 2}

Here we evaluate the generalized factorial moments for variables related
to transverse momentum, given by (\ref{t16}), in the limit of very large
total number of produced particles (but without any restriction on the
ratio $\hat{n}/\hat{N}$, i.e. on the size of the $s$-bin). 

The starting point is the formula .

Developping the function $\phi(y)$ around $y=0$ we have
\ba
\phi(y) = \sum_{m=0}^\infty \phi_m \frac{(iy)^m}{m!}  \label{a21}
\ea
where
\ba
\phi_m =\frac1{2\pi D^2}\int d^2q x(q) q^m e^{-q^2/2D^2}= \overline{x(q)
q^m} \label{phm}
\ea
This implies that $[\phi(y)]^k$ can also be represented by a series
\ba
[\phi(y)]^k = \sum_{m=0}^\infty \phi_m^{(k)}y^m
\ea
where the coefficients $\phi_m^{(k)}$ can be explicitely evaluated, when
necessary.

Introducing this in (\ref{t16}) we have
\ba
\frac{S_{k\perp}}{S_{0\perp}}\approx  \hat{n}^k\sum_{m=0}^\infty
\phi_m^{(k)}
\left<y^{2m}\right>/m!= \hat{n}^k\sum_{p=0}^\infty
p!\phi_{2p}^{(k)}\left(\frac2{\hat{N}\Delta^2}\right)^{p}
  \label{t162}
\ea
One sees that in the limit  of very large $\hat{N}$ the first 
nonvanishing term dominates. If $\phi_0= \bar{x}\neq 0$ we have
\ba
F_{k\perp}[x(q)]\approx  \left<\hat{n}^k \phi_0^{(k)}\right>= 
\left<[\hat{n}\bar{x}]^k\right>  
  \label{t16y}
\ea
 and thus we recover the original formula (\ref{p9}). We conclude that in
this case  the standard interpretation of the generalized factorial moments
 remains valid even in presence of the conservation law.

If $\bar{x}=0$,  and $\phi_1=\overline{qx} \neq 0$, we have 
\ba
[\phi(y)]^k =
[iy\phi_1]^k\left[1+iy\frac{k\phi_2}{2\phi_1} +...\right]=
[iy\phi_1]^k + (iy)^{k+1}\phi_1^{k-1}\phi_2/2+....
\ea
For $k$ even, $k=2p$ the dominant term is
\ba
\frac{S_{k\perp}}{S_{0\perp}}\approx 
 (-1)^p \hat{n}^{2p} [\phi_1]^{2p} \frac{\hat{N}\Delta^2}{2\pi}
\int d^2y (y^2)^pe^{-y^2\hat{N}\Delta^2/2}
= \nn=
 (-1)^p [\hat{n}\overline{qx}]^p p! \left(\frac{2\hat{n}\overline{qx}}
{\hat{N}\Delta^2}\right)^p
  \label{t16x}
\ea
where we have used the relation (\ref{phm}).

For $k$ odd, $k=2p-1$ we obtain

\ba
\frac{S_{k\perp}}{S_{0\perp}}\approx (-1)^p\frac12\hat{n}
^{2p-1}[\overline{qx}]^{2p-2}\overline{q^2x}
\frac{\hat{N}\Delta^2}{2\pi}\int d^2y (y^2)^p
e^{-y^2\hat{N}\Delta^2/2 }=\nn=
(-1)^p\frac12\hat{n}
^{p-1}[\overline{qx}]^{p-2}\overline{q^2x}
p! \left(\frac{2\hat{n}\overline{qx}}
{\hat{N}\Delta^2}\right)^p.  \label{z}
\ea
where again (\ref{phm}) was used. Note that 
 these moments are proportional to $\overline{q^2x}$ and thus vanish if the variable
$x(q)$ is odd with respect to change $(q\leftrightarrow -q)$.

We conclude that the standard interpretation of the factorial moments,
as expressed in (\ref{p9}), holds only when the average value  $\overline{x}$
does not vanish. When $\overline{x}=0$ the factorial moments measure the 
average value $\overline{qx}$ rather than $\overline{x}$.

\vspace{0.3cm}
{\bf Acknowledgements}
\vspace{0.3cm}

I would like to thank Andrzej Kotanski for pointing out an inconsistency
in the original formulation \cite{gfm} of the idea presented in this
paper. I am also grateful to Krzysztof Fialkowski for a critical
reading of the manuscript. This investigation was supported in part by
the MEiN Grant No 1 P03 B 04529 (2005-2008).


\begin{thebibliography} {99}
\bibitem{bk}
A.Bialas and V.Koch, Phys.Lett. B456 (1999) 1.
\bibitem{gfm}
A.Bialas, Acta Phys. Pol. B35 (2004) 683.
\bibitem{vs} 
S. Voloshin and D.Seibert, Phys. Lett. B249 (1990) 321.
\bibitem{sv}
  D.Seibert and S. Voloshin, Phys. Rev. D43 (1991) 119; 
D.Seibert, Phys. Rev. C44 (1991) 1223. 
\bibitem {vkr}
S.Voloshin, V.Koch and H. Ritter, Phys. Rev. C60 (1999) 024901.
\bibitem{pgv}
C.Pruneau, S.Gavin and S.Voloshin, Phys. Rev.  C66 (2002) 044904; Nucl.Phys. A715 (2003) 661.
\bibitem{ag}
S.Gavin, Phys. Rev. Lett. 92 (2004) 162301; M.Abdel-Aziz and S.Gavin,
Mucl. Phys. A774 (2006) 623.
\bibitem{jk}
S.Jeon and V.Koch, in "Quark-Gluon Plasma 3", World Scientific (Singapore
2003), p.430.
\bibitem{v}
S.Voloshin, nucl-th/0206052.
\bibitem {bp1}
A.Bialas and R.Peschanski, Nucl. Phys. B273 (1986) 703;
 Nucl. Phys. B308 (1988) 857.
\bibitem {gm}
M.Gazdzicki and S.Mrowczynski, Z.Phys., C54 (1992) 127.
\bibitem{na49}
H.Appelshauser et al., NA49 coll., Phys. Lett. B459 (1999) 679.
\bibitem{phenix}
K.Adkox et al., PHENIX coll., Phys. Rev. C66 (2002) 024901.
\bibitem{ceres}
D.Adamova et al, CERES coll., Nucl. Phys. A727 (2003) 97.
\bibitem{star}
J.Adams, et al., STAR coll., Phys. Rev. C72 (2005) 044902.
\bibitem{na22}
M.Atayan et al., NA222/EHs coll., Phys. Rev. D71 (2005) 012002.
\bibitem {fbhb}
W.Broniowski, B.Hiller, W.Florkowski and P.Bozek, Phys. Lett. B635 (2006) 290; W.Florkowski, W.Broniowski, B.Hiller and P. Bozek, nucl-th/0610035.
\bibitem {fl} 
J.Fu, L.Liu, Phys. Rev. C68 (2003) 064904.
\bibitem{jyj}
J. Fu, Y.Gao and J.Cheng, Phys. Rev. C72 (2005) 017901.
\bibitem{o}
G.Odyniec, Acta Phys. Polon. B30 (1999) 385.
\bibitem{bdo}
N.Borghini, P.M.Dingh, J.-Y. Ollitrault, Phys. Rev. C62 (2000) 034902.

\end{thebibliography}
\end{document}